\documentclass{wiley-article}
\usepackage[numbers]{natbib}
\usepackage{siunitx}
\usepackage{lipsum}
\usepackage{amsfonts}
\usepackage{graphicx}
\usepackage{epstopdf}
\usepackage{algorithmic}
\usepackage{blindtext}
\usepackage{multicol,multirow}
\usepackage{float,caption}
\usepackage{hyperref}
\usepackage{graphicx}%,dblfloatfix}
\usepackage{amsmath}
\usepackage{fancyhdr}

\newcommand{\pard}[2]{\frac{\partial #1}{\partial #2}}

\papertype{Original Article}
\paperfield{Journal Section}

\title{A One Dimensional (1D) Computational Fluid Dynamics Study of Fontan-Associated Liver Disease (FALD)}

\author[1]{Yaqi Li}
\author[2]{Justin D. Weigand, M.D.}
\author[3]{Charles Puelz, Ph.D.}
\author[4]{Mette S. Olufsen, Ph.D.}
\author[5]{Alyssa Taylor-Lapole, Ph.D.}

\affil[1]{Department of Brain and Cognitive Sciences, Massachusetts Institute of Technology, Cambridge, MA, 02139, USA}
\affil[2]{Division of Cardiology, Department of Pediatrics, Baylor College of Medicine and Texas Children's Hospital, Houston, TX, 77030, USA}
\affil[3]{Mathematics, University of Houston, Houston, TX, 77204, USA}
\affil[4]{Department of Mathematics, North Carolina State University, Raleigh, NC, 27695, USA}
\affil[5]{Computational Applied Mathematics and Operations Research, Rice University, Houston, TX, 77251, USA}

\corraddress{Alyssa Taylor-LaPole, Computational Applied Mathematics and Operations Research, Rice University, Houston, TX, 77251, USA}
\corremail{al207@rice.edu}

\fundinginfo{NSF, Grant/Award Numbers: DGE-2137100, DMS-2051010. NIH, Grant/Award Number: 1L40HL181846-01.}

\runningauthor{Yaqi Li et al.}

\begin{document}

\begin{frontmatter}
\maketitle

\begin{abstract}
Fontan-Associated Liver Disease (FALD) is a disorder arising from hemodynamic changes and venous congestion in the liver. This disease is prominent in patients with hypoplastic left heart syndrome (HLHS). Although HLHS patients typically survive into adulthood, they have reduced cardiac output due to their univentricular physiology (i.e., a Fontan circuit). As a result, they have insufficient blood delivery to the liver. In comparison, patients with double outlet right ventricle (DORV), also having a univentricular circuit, have lower incidence of FALD. In this study, we use a patient-specific, one-dimensional computational fluid dynamics (1D-CFD) model to predict hemodynamics in the liver of an HLHS patient and compare predictions with an age- and size-matched single-ventricle Fontan DORV control patient. Additionally, we simulate FALD conditions in the HLHS patient to predict hemodynamic changes across various stages of disease progression. Our results show that the HLHS patient higher hepatic arterial pressure compared to the DORV patient. This difference is exacerbated as FALD conditions progress. HLHS patients also have higher average portal pressures than the DORV. The wall shear stress (WSS) is higher in the hepatic network for the simulated FALD patients. WSS is slightly decreased in the portal network for the HLHS patients, consistent with the development of portal hypertension. Perfusion analysis gives insight into regions of liver tissue at risk for fibrosis development, showing increasing pressures and reduced flow throughout the liver tissue fed by the portal vein under FALD conditions. Our results provide insight into the specific hemodynamic changes in Fontan circulation that can cause FALD.
% Please include a maximum of seven keywords
\end{abstract}
\end{frontmatter}

\section{Introduction}

\noindent The Fontan procedure is a life-saving surgery for single ventricle patients. Single ventricle disease is a category of rare congenital heart defects that affect about 5 in 100,000 babies \cite{Fixler2010, Reller2008}. The most common type is hypoplastic left heart syndrome (HLHS). HLHS patients are born with an underdeveloped left ventricle and aorta, leaving them with no way to pump oxygenated blood sufficiently to the body \cite{noonan1958hypoplastic}. Patients undergo a series of surgeries from their first weeks of life to age three, resulting in a fully functioning univentricular circulatory system \cite{tellez2018fontan}. This system is called the Fontan circuit. For these patients, the  single functioning right ventricle supports the pulmonary and systemic circulations; the ventricle actively pumps blood into the systemic circulation, while the pulmonary circulation is driven by a passive pressure gradient from the systemic veins to the lungs \cite{fontan1971surgicalrepair}. This leads to an elevation in central venous pressure (CVP) \cite{leeshahjehan2023fontancompletion}. Approximately $61\%–85\%$  of patients with Fontan circuits survive until early adulthood, but these patients will suffer long-term complications.

A common complication is the development of Fontan-associated liver disease (FALD) \cite{khairy2008long, downing2017long,schilling2016fontan, kverneland2018five}. Almost {all} Fontan patients will develop FALD to some degree, which could lead to portal hypertension, hepatocellular carcinoma, and eventual liver failure \cite{emamaullee2020fontan}. Due to these complications, patients require liver and/or heart transplants around the age of 20 \cite{vaikunth2019short, deLange2023}. 

{Blood enters the liver through two conduits and is drained by the hepatic veins, making it a complex system to model.} The hepatic artery, branching from the abdominal aorta, transports $25-30\%$ of blood to the liver, while the portal vein supplies the other $70-75\%$ \cite{lautt2009hepatic, eipel2010regulation, rocha2012liver}. The portal vein and hepatic artery systems have a significant pressure difference.  The hepatic artery has a mean pressure of $90$ mmHg, and the portal vein has a mean pressure of $10$ mmHg \cite{granger2004circulation}. Studies show that increased pressures of the portal vein or hepatic artery lead to liver damage such as cirrhosis, fibrosis, nonalcoholic fatty liver disease, and portal hypertension \cite{xanthopoulos2019heart, alvarez2011liver}. Understanding the flow and pressure within the liver vasculature due to a Fontan circulation is vital to understanding the pathology of FALD and predicting its development.

FALD results from an increase in the venous pressure in the portal vein caused by an increase in the central venous pressure (CVP) typical of this group of patients. Despite the increased pressure in the portal vein, the pressure gradient is reduced, which reduces blood flow to the liver. The result is chronic systemic venous congestion and reduced cardiac output \cite{navaratnam2016exercise,vandebruaene2015effect,oka2023liver,mazza2021pathophysiology}. 

FALD progresses through several stages before reaching a final, irreversible stage, where the main complications and the breakdown of the circulatory system occur \cite{gordonwalker2019fontanliver}. The initial stages of FALD are characterized by liver congestion and sinusoidal dilation. For most patients, this begins at birth and continues after Fontan surgery \cite{goldberg2017hepaticfibrosis, johnson2013identifying, schwartz2012hepatic}. Although many patients remain without symptoms, $53\%$ eventually experience painful hepatomegaly or hepatojugular reflux due to reduced cardiac output \cite{tellez2018fontan}. The second stage of FALD involves perisinusoidal fibrosis, regenerative nodules, and hepatocellular necrosis, which occurs 5-10 years after surgery \cite{tellez2018fontan}. In most patients, the final stage involves advanced fibrosis with portal vein hypertension, leading to inadequate filtration of the blood within the liver \cite{rychik2002relation, tellez2018fontan}. As FALD progresses, these complications lead to increased stiffening of the vasculature \cite{Daniels2014}.

Currently, the underlying pathology and physiology of FALD are not fully understood. Most FALD studies are based on animal models or clinical data from patient cohorts with other chronic liver diseases \cite{delange2023fontanliverdisease}. Few studies focus on the pathology of FALD. A review by Cieplucha \textit{et al.} \cite{cieplucha2022fontanliverdisease} suggests that patients with HLHS develop FALD due to two factors: 1) permanent reduction in cardiac output and altered hemodynamics due to Fontan reconstruction, and 2) {reduced oxygen delivery resulting from reduced cardiac output. It has been hypothesized that the reconstructed aorta of HLHS Fontan patients further reduces cardiac output compared to other Fontan patients without reconstruction. This is due to abnormal aortic flow leading to power losses and stagnant flow in the aorta \cite{sundareswaran2006} and increased aortic stiffness over time \cite{Walser,Rychik}}. Several clinical studies suggest that passive flow in the Fontan circulation to the pulmonary system is analogous to a cardiovascular system with obstruction in the hepatic venous outflow \cite{delange2023fontanliverdisease, schleiger2020evaluation}. In a report by \cite{Daniels2014}, elastographies of Fontan patients revealed that the longer a patient lives with a Fontan circuit, the stiffer their liver vasculature will be. However, it is unknown whether the increase in stiffness is due to increased congestion (i.e., elevated pressures in the liver) or the development of fibrosis. 

Computational studies of the Fontan circulation are limited. Most computational studies use three-dimensional computational fluid dynamics (3D-CFD) to investigate velocity patterns and power dissipation \cite{marsden2007effectsofexercise, ahmed2021interventionalplanning}. While this technique allows the assessment of complex velocity patterns, it is time-consuming and unrealistic for large-body network models. Moreover, most of these studies focus on the Fontan reconstruction rather than the effects of a Fontan circuit on downstream organs.  

An alternative is one-dimensional fluid dynamics (1D-CFD) models. These models efficiently forecast wave propagation within extensive networks and can encompass fluid-solid interactions. For example, the study by \cite{puelz2017computationalstudy} uses a 1D arterial and venous network with zero-dimensional heart and organ bed models to study variations in Fontan circulations, contrasting Fontan circuits with fenestration versus hepatic vein exclusion. This study indicates enhanced gut flow but did not consider the potential influence of vortices in bifurcations and reconstructed vessel segments. A recent study by Taylor-LaPole \textit{et al.} \cite{alyssa2022} uses a patient-specific 1D-CFD network to predict the effect of aortic reconstruction in {two single ventricle Fontan patients: an HLHS patient and a double outlet right ventricle (DORV) patient. The latter, which serves as the control, has the same physiology as the HLHS patient, but does not have a reconstructed aorta. DORV is a congenital heart defect in which the pulmonary artery and aorta connect to the right ventricle; many of these patients also have a nonfunctioning left ventricle \cite{goo2021double}.} This study finds that the HLHS patient has hypertensive pressures in the brain and reduced flow to the gut. These studies present promising large vessel networks of the Fontan circulation, but explore neither organ-specific networks nor the potential development of FALD.

Computational studies focusing specifically on liver disease in the Fontan circuit are scarce. The study by Trusty \textit{et al.} \cite{trusty2018impact} uses computational fluid dynamics simulations to quantify the total cavo-pulmonary connection and cross analyze to data from quantified liver biopsies. Although this helps determine effects within the constructed Fontan conduit downstream of the liver, it does not provide detailed information about the liver vasculature itself. The study by Ishizaki \textit{et al.} \cite{ishizaki2018prediction} develops a vortex flow map to predict blood movement within the right atrium during the cardiac cycle. The study suggests that reduced vortex flow in the right atrium during the late phase of the Fontan operation is correlated with the development of FALD but does not include liver circulation in the model itself.

\begin{figure}[H]\centering
\includegraphics[width=0.7\linewidth]{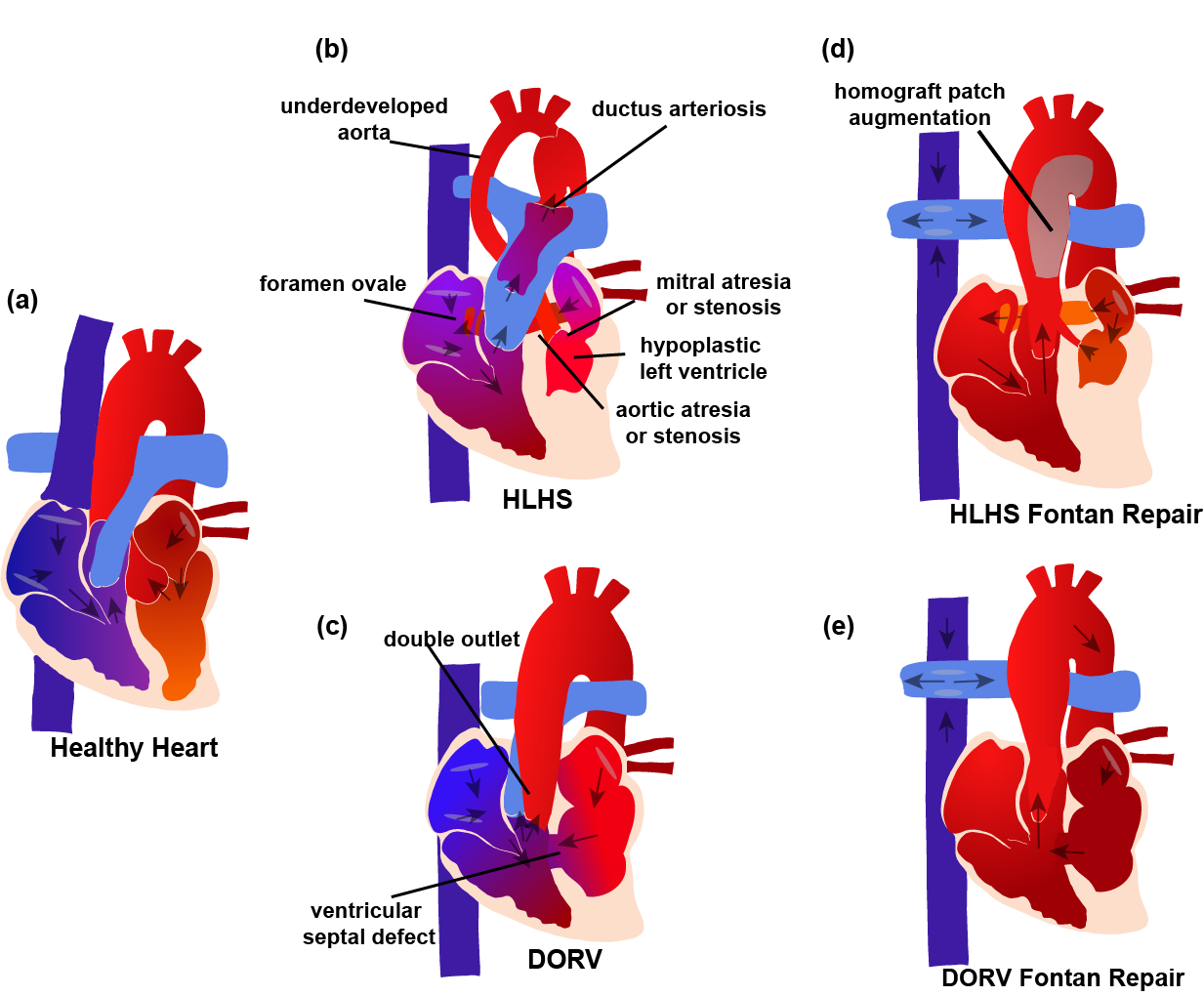}
\caption{{(a) healthy heart with two ventricles. (b) hypoplastic left heart syndrome (HLHS) heart. (c) double outlet right ventricle (DORV) heart with a ventricular septal defect. (d) HLHS heart with a Fontan repair. This patient has undergone aortic reconstruction. (e) DORV heart with a Fontan repair. Patients in (d) and (e) both have a total cavopulmonary connection with the main pulmonary artery connected to the inferior vena cava.}}
\label{fig:circulations}
\end{figure}

To our knowledge, no previous computational studies have analyzed the effect of aortic reconstruction on FALD hemodynamics while incorporating a detailed network of liver vasculature. To investigate the effects of a reconstructed aorta on the development of FALD, we compare {the two single ventricle patients with Fontan circulations from \cite{alyssa2022}: an HLHS patient and a DORV patient. Both of these patients have a Fontan circuit, but the DORV patient did not have a surgical reconstruction of the aorta \cite{goo2021double}. Figure \ref{fig:circulations} shows the two patients' along with the connectivity of a healthy control with a two-sided heart.} 

To investigate the progression of FALD in a Fontan circuit, we impose increased vascular stiffness and resistance on the patient with HLHS, mimicking the effects of increased CVP. We compare liver perfusion between the patient with HLHS without FALD, the control patient with DORV, and the patients with HLHS and varying degrees of FALD. This is done using a 1D-CFD model predicting pressure, flow, and wall shear stress (WSS) in patients with HLHS and DORV. Our detailed liver network model allows us to investigate pressure changes and blood perfusion within the liver, providing new insight into hemodynamics in the liver of a Fontan circuit, including both arterial and portal vein circulations.  {We investigate patient-specific differences by comparing hemodynamic responses with normalized and patient specific cardiac outputs.}

\section{Methods}
This study uses hemodynamic data from a previous study \cite{alyssa2022} along with computed tomography (CT) images of the liver to construct an arterial network of both the portal and hepatic vasculature. The liver image is from a representative adult male obtained from an open-access repository. Therefore, vascular dimensions were allometrically scaled as done in \cite{alyssa2022} to match the specific patients in this study. A 1D CFD model is solved within these networks to predict pressure and flow. The solution is mapped to the liver tissue providing insight into tissue perfusion and distributions of pressure. FALD is simulated in patients with aortic reconstruction by increasing downstream resistance, thus inducing increased CVP. 
\begin{figure}
\centering
\includegraphics[width=1.0\linewidth]{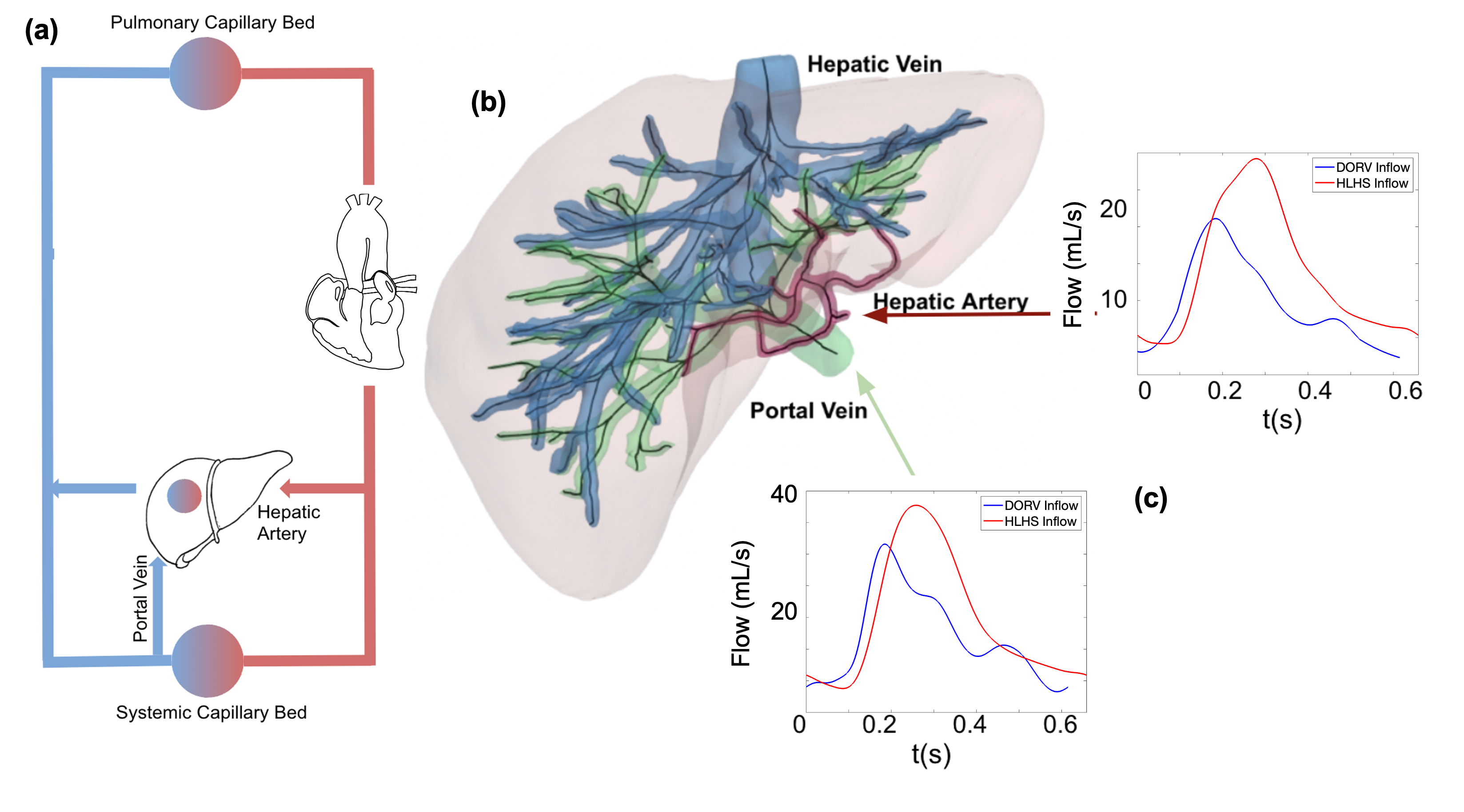}
\caption{ {(a) The Fontan circulation, including flow to the liver. The right heart pumps blood into the aorta, which feed the liver via the portal vein, supplying 70\% of the blood, and the hepatic artery, supplying 30\% of the blood \cite{rocha2012liver}. From the liver, blood is drained into the  systemic veins, which passively is passing through the pulmonary circuit returning to the right heart. (b) 3D rendering of the hepatic artery (blue-green) and portal vein (red) vasculatures extracted from a CT scan from a healthy adult. Centerlines extracted from VMTK are shown in black. The network generated is scaled to the height and weight of the two subjects analyzed here. (c) Inflow waveforms (mL/s) the hepatic artery and portal vein for the two patients. The DORV patient has a greater occurrence of sub-oscillations, that is reflective waves after the initial peak, at the inlet of both networks.}}
\label{fig:centerlines}
\end{figure}

\subsection{Data and Image analysis}
\label{sec:segment}
%This study includes simulations of two representative single ventricle subjects with Fontan circulation (Figure \ref{fig:circulations}): a DORV and HLHS subjects. To facilitate the comparison of the outcomes associated with aortic remodeling, subjects were generated as a pair. As noted in Table 1 they have similar weight, height, age, cardiac output, and heart rate. The main difference is that the subject with HLHS had aortic reconstruction but did not, and their blood pressure also differs.

This study includes simulations of two representative single ventricle subjects with Fontan circulation (Figure \ref{fig:circulations}): a DORV and an HLHS subject. Subjects were paired, matching the outcomes associated with aortic remodeling. Table \ref{Tab:strokeVolume} shows they have similar weight, height, age, blood pressure, and heart rate, but their cardiac output differs. The HLHS patient has a higher cardiac output than the DORV patient.

 {The DORV and HLHS patients used in this study were part of a larger pilot study using three-dimensional, time-resolved phase contrast cardiac magnetic resonance (CMR) to assess differences in aortic flow properties between functional single right ventricle with and without aortic arch reconstruction. Single ventricle patients who had arch reconstruction (HLHS) and age/sized matched patients with native aortic arches (DORV) with mitral atresia were prospectively recruited.  All subjects had a single right ventricle anatomy, less than moderate atrioventricular valve regurgitation, and a total cavopulmonary anastomosis (non-fenestrated). Patients with aortic surgery beyond the initial Norwood and significant collateral burden (<40\% "Qp" ), systemic hypertension, and those with heart failure were excluded. Thus, patients, included in this study, had no evidence of failing Fontan physiology by means of cardiac dysfunction or AV valve failure.}

A CT image of the liver region of an adult male volunteer with tumors in the pancreas is used to construct a 3D representation of the vasculature \cite{soler2010}. We assume that the liver vasculature is normal and is not affected by the tumor. The image can be found in the publicly available 3D-IRCADb-01 database (see \href{https://www.ircad.fr/research/data-sets/liver-segmentation-3d-ircadb-01/}{https://www.ircad.fr/research/data-sets/liver-segmentation-3d-ircadb-01/}). From this image, we extract vascular networks distal to the hepatic artery and the portal vein.

Segmentations of the hepatic artery and portal vein networks are generated using the open-source software 3D Slicer developed by Kitware, Inc. (see \href{http://www.slicer.org}{http://www.slicer.org}) \cite{Federov12, Kikinis14}. Using built-in tools from 3D Slicer, including painting, thresholding, cutting, and islanding, a 3D rendering of the arterial and portal vein networks is obtained. The image intensities for the hepatic artery and portal vein segmentations are set to 28.05-255.00 {Hounsfield} units. These segmentations are exported as STL files from 3D Slicer and imported to Paraview \cite{Utkarsh15}, where they are converted to a VTK polygonal data file for further processing. The liver volume is also segmented. The volume is segmented in the same fashion as the vasculature, with image intensity set to 64.00-255.00 {Hounsfield} units. The resulting segmentation is exported as an STL file and opened in Paraview, where it is converted to a volumetric mesh and saved as a csv file storing the $x,y,z$ coordinates of the tissue. 

We utilize the Vascular Modeling Toolkit (VMTK) to generate centerlines in 3D renderings of the hepatic artery and portal vein network \cite{Antiga2008}. This software places maximum inscribed spheres at each $x,y,z$ coordinate of the vessels and approximates the medial axis with a Voronoi diagram \cite{Antiga2008,izzo2018vascular}. Centerlines are determined as the minimal paths along the inverse of the spheres' radii \cite{Antiga2008}. Inlet and outlet boundaries are manually set, and VMTK works recursively from the terminal to the proximal vessel(s), generating centerlines. 

\noindent\textbf{Labeled Directed Graph}
We use in-house algorithms in MATLAB to extract network information from VMTK and construct a labeled directed tree with edges defining vessels, nodes, and junctions. Each vessel has labels that denote its radius and length \cite{Colebank19,Colebank21}. VMTK defines the points of centerlines as $x$, $y$, and $z$ coordinates along each vessel. The locations at which two central lines interact are junctions. They are adjusted to appear in the ostium's center using the junction correction algorithm by Bartolo \& Taylor-LaPole \textit{et al.} \cite{bartolo2023framework}. The vessel and its labels are integrated into a directed tree, defining a connectivity matrix that connects the vessel to its parent and daughters. Optimal vessel dimensions, i.e., radii and length, are determined using an in-house algorithm utilizing statistical changepoints \cite{bartolo2023framework}.

\begin{figure}
    \centering
    \includegraphics[width=0.75\linewidth]{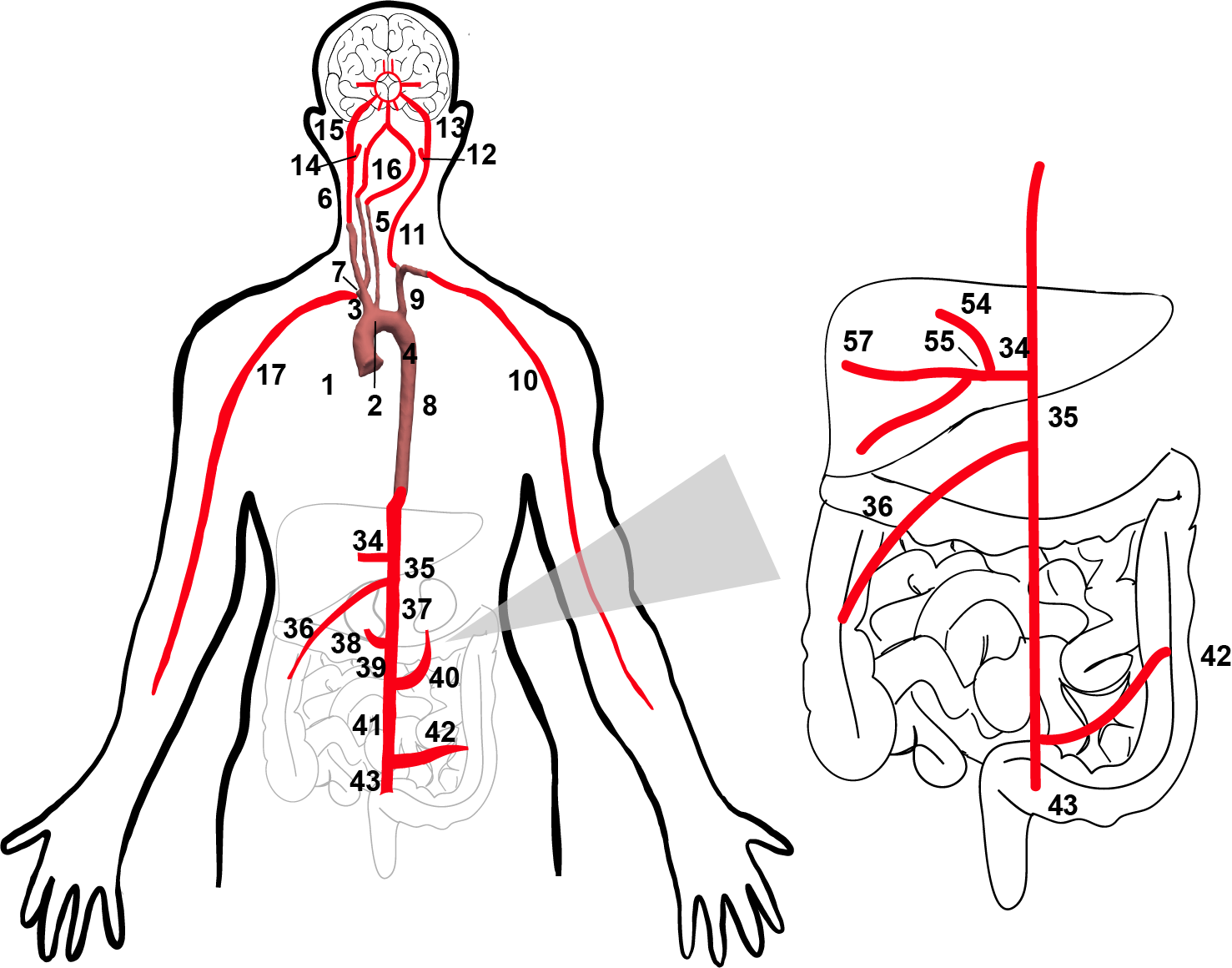}
    \caption{Network adapted from \cite{alyssa2022} used to calculate the inflow to the liver. Flow in vessel 34 feeds the hepatic artery, and flow from vessels 36, 40, and 42 enters the portal vein.}
    \label{fig:network}
\end{figure}

\noindent\textbf{Hemodynamics Data}
{Approximately 25\% of cardiac output enters the liver \cite{lautt2009hepatic, rocha2012liver}. The inflow to the liver is determined by predicting the flow to the hepatic artery and portal vein using the calibrated whole-body model from \cite{alyssa2022}. For both patients, the hepatic artery flow is taken from predictions in vessel 34 (Figure 3) and the flow to the portal vein is obtained by combining flows through the intestinal vasculature (vessels 36, 40, and 42). The DORV patient has a total flow to the liver of 1.12 L/min (28$\%$ of cardiac output), and the HLHS patient has a liver flow of 1.30 L/min (26$\%$ of cardiac output). Values are listed in  Table \ref{Tab:strokeVolume} the two patients studied here have similar cardiac output. The differences in aortic vasculature result in the patient with DORV having a total flow to the liver of 1.12 L/min and the patient with HLHS having an flow of 1.30 L/min (listed in Table \ref{Tab:strokeVolume}). Of these, approximately 70$\%$ enter through the portal vein and 30$\%$ through the hepatic artery \cite{jamshidi}. Again, these distributions agree with our findings; For the DORV patient, the flow split is (37$\%$, 63$\%$) and for the HLHS patient, it is (34$\%$, 66$\%$). It should be noted that the flow through the portal vein does not reflect the path through the intestines. Therefore, the pulsatility of the wave is likely too high. It is well known that the flow waveform through the portal vein varies significantly between subjects and is likely blunted in patients with a single ventricle \cite{chavan}. To compare results with a pulsatile (shown in Figure 2) versus a blunted (shown in supplementary material Figure 1) waveform, we performed simulations with the waveform obtained by combining flows in vessels 36, 40, 42 and scaling a digitized and smoothed waveform from \cite{chavan}.}

%The inflow to the liver is determined by scaling the measured cardiac output \cite{alyssa2022} to this amount. The DORV patient has a total hepatic inflow of 1.05 L/min and the HLHS patient has an inflow of 1.27 L/min. Of these, 70\% enters through the portal vein and 30\% through the hepatic artery. Figure \ref{fig:centerlines} shows the scaled inflow waveforms, and Table \ref{Tab:strokeVolume} lists the flows for each patient. 
\begin{table}[t]
\centering
 \caption{Patient characteristics, cardiac output and flow to the liver. Patient characteristics and cardiac output are taken from \cite{alyssa2022}.  {*The cardiac outputs are for the DORV and HLHS patients described in section 2.1. For simulations in section 2.5, the HLHS cardiac output was imposed on the DORV network (simulation 5) and the DORV cardiac output was imposed on the DORV network (simulation 6.}}
\begin{tabular}{lll}

\hline
                      & DORV & HLHS \\ 
\hline
Age (years) & 12 & 11   \\
Height (cm) & 154.3 & 151.4 \\
Weight (kg) & 59.6 & 62.0 \\
Blood pressure (mmHg) & 110/67 & 116/65 \\
Cardiac output (L/min)* & 4.06 & 5.08 \\
Total liver CO (L/min)* & 1.12 & 1.30 \\
Portal vein flow (L/min)* & 0.71 & 0.86  \\
Hepatic artery flow (L/min)* & 0.41 & 0.44 \\ 
\hline
\end{tabular}
  \label{Tab:strokeVolume}
\end{table}

\subsection{One Dimensional Fluid Dynamics Model}
Similarly to \cite{Colebank19,alyssa2022, olufsen2000numerical,bartolo2023framework}, we predict pressure $p(x,t)$ (mmHg), blood flow $q(x,t)$ (mL/s), and area deformation $A(x,t)$ (cm$^2$) in each vessel within the portal vein and hepatic artery vasculatures using a 1D fluid dynamics model. The model is derived from the Navier-Stokes equations, which assume that blood flow is axisymmetric and Newtonian. We also assume that blood is incompressible, viscous, and homogeneous with constant density $\rho$, and that each vessel can be represented by a cylinder with an impermeable wall. Under these assumptions, mass conservation and momentum balance are governed by
\begin{align}
\frac{\partial A}{\partial t} + \frac{\partial q}{\partial x} &= 0 \label{eq:mass} \\
\frac{\partial q}{\partial t} + \frac{\partial}{\partial x}\left(\frac{q^2}{A}\right) + \frac{A}{\rho}\frac{\partial p}{\partial x} &= -\frac{2\pi \nu R}{\delta} \frac{q}{A}, \label{eq:mom}
\end{align}
where $0 \leq x \leq L$ is the axial position within the vessel, $R$ (cm) is the radius, $\mu$ (g/cm/s) is the viscosity, and $\nu = \frac{\mu}{\rho}$ (cm$^2$/s) is the kinematic viscosity. 

{In the organ network studied here, the vessels are relatively short compared to their length. As a result, we do not anticipate a fully developed flow. To account for this, we impose a Stokes velocity profile} with a linearly decreasing boundary layer with thickness $\delta = \sqrt{\nu T/2\pi}$ (cm) \cite{nichols1991mcdonald,pedersen1993two,pries1992blood,olufsen2000numerical}, where $T$ (s) is the length of the cardiac cycle.
\begin{align}
    u_x(r,x,t) = \begin{cases}
    \bar{u}_x, \ \ \ r<R-\delta\\
    \frac{\bar{u}_x}{\delta} (R-r), \ \ \ R-\delta < r\leq R.
    \end{cases}
    \label{stokes}
\end{align}
In this equation, $u_x$ denotes the axial velocity and $\bar{u}_x$ is the axial velocity at the center of the vessel. This system of equations is closed via a pressure-area relationship modeled as a linear elastic membrane
\begin{equation}
    p(x,t)-p_0 = \frac43 \frac{Eh}{r_0}\left(1-\sqrt{\frac{A_0}{A}}\right), \hspace{1cm}
    \frac{Eh}{r_0} = k_1 e^{k_2r_0}+k_3, \label{eq:ehr0}
\end{equation}
where $E$ (g/cm/s²) is Young's modulus, $h$ (cm) is the vessel wall thickness, $p_0$ (mmHg) is the reference pressure, $r_0$ (cm) is the inlet radius, and $A_0$ (cm²) is the cross-sectional area when the pressure equals its reference value \cite{qureshi2019hemodynamic}. We assume that the vessels become stiffer as they get smaller (equation (\ref{eq:ehr0}), where $k_1$ (g/cm/s$^2$), $k_2$ (1/cm), and $k_3$ (g/cm/s$^2$) are constants obtained from literature values \cite{Colebank21,paun2020assessing,alyssa2022}.

The system is hyperbolic, so boundary conditions are required at the inlet and outlet of each vessel. At the inlet of the portal vein and hepatic artery, we assign a waveform as described above. At the junctions, we impose mass conservation and pressure continuity by	
\begin{align}
    p_p(L, t)  &= p_{d_i}\\
    q_p(L, t)  &= \sum q_{d_i}.
\end{align}
where $p$ is a parent vessel and $d_1$ and $d_2$ are daughter vessels. Similarly to previous studies \cite{Colebank19, alyssa2022, olufsen2000numerical, bartolo2023framework}, we solve the model using the two-step Lax-Wendroff finite difference scheme. 

At the outlet of the terminal vessels, we use a structured tree boundary condition prescribing flow to the small vessels in the vascular bed \cite{colebank2021sensitivity,  olufsen2000numerical,  alyssa2022,  chambers2020structural}. Equations in the structured tree are obtained semianalytically by solving a linearized wave equation ignoring inertial forces, which is appropriate since viscous forces are dominant in small vessels.  

The resulting wave equations stated in the frequency domain, described in detail in \cite{olufsen2000numerical}, are given by
\begin{align}
    \frac{\omega^2}{c}Q+\pard{^2Q}{x^2}=0, \hspace{1cm} c=\sqrt{\frac{A_0(1-F_J)}{\rho C}}, \label{eq:SmallVes3}
\end{align}
where $c = f(F_J, C)$ denotes the wave propagation velocity \cite{olufsen2000numerical}. The function $F_J(J_0, J_1)$ is a function of $J_0$ and $J_1$, the zeroth- and first-order Bessel functions, and $C$ is the vessel compliance. Solving the equation for flow $Q$ and pressure $P$, we get
\begin{align}
    Q(x,\omega) &= a\cos(\omega x/c)+b\sin(\omega x/c)\\
    P(x,\omega) &= i\sqrt{\frac{\rho}{CA_0(1-F_J)}}
    (-a\sin(\omega x/c)+b\cos(\omega x/c)).\label{eq:SmallVesSolution}
\end{align}
In this equation $a$ and $b$ are integration constants \cite{alyssa2022,Colebank21, olufsen2000numerical, bartolo2023framework}.

Similar to large vessels, we enforce pressure continuity and flow conservation at junctions as
\begin{align}
    \frac{1}{Z_p}=\frac{1}{Z_{d_1}}+\frac{1}{Z_{d_2}},
\end{align} 
where $Z_p, Z_{d_1}$, and $Z_{d_2}$ are the impedance values of the parent and daughter vessels. {It is assumed that the terminal impedance at the end of the structured tree is zero.} The equations in this tree are solved recursively for each vessel, predicting the impedance at the beginning of the vessel as a function of the impedance at the root of the vessel. The impedance at the root of the structured tree is transformed to the time domain by convolution. Following Riemann invariants, the root impedance forms the outflow boundary condition for large vessels.

\subsection{Wall Shear Stress Approximation}
Wall shear stress (WSS), the stress the fluid exerts on the vessel wall, is denoted by $\tau_w$ (g/cm/s²). With results from flow and area predictions, we approximate the WSS for selected vessels in the portal vein and hepatic artery vasculatures (vessel numbers 0, 54, 58, 69 in the portal vein vasculature and 0, 13 in the hepatic artery vasculature) using the Stokes boundary layer given in Equation \ref{stokes} with
\begin{equation}
\tau = -\mu \frac{\partial u}{\partial r}
\end{equation}
\begin{equation}
\tau_w=
\left\{
\begin{array}{lr}
0, & r < R - \delta\\
\frac{\mu\bar{u}}{\delta}, &R - %\delta < r \leq R\\
\end{array} \right.
\end{equation}
where $\mu$ is blood viscosity, and $\delta = \sqrt{\frac{\nu T}{2\pi}}$ is the boundary layer thickness \cite{alyssa2022, bartolo2023framework}.
\begin{table}[t]
\centering
\caption{Patient-specific parameter values. PV refers to the portal vein, HA to the hepatic artery. {$k_{1,2,3}$ denote large vessel stiffness, $ks_{1,2,3}$ denote small vessel stiffness, $p_0$ is the diastolic pressure coefficient, $T$ is length of one cardiac cycle, and $r_{\text{min}}$ denotes the terminal radius value that the structured tree is propagated to.}}
\begin{tabular}{lrrrrrrrr}
\hline
\multicolumn{1}{c}{\multirow{2}{*}{Parameter}} & \multicolumn{2}{c}{Healthy HLHS} & \multicolumn{2}{c}{Early FALD} & \multicolumn{2}{c}{Late FALD} & \multicolumn{2}{c}{DORV} \\
\multicolumn{1}{c}{}  & PV  & HA & PV & HA & PV & HA & PV & HA         \\ \hline
$k_1$ (g/cm/s$^2$) & 2e+6 & 2e+6  & 2e+6 & 2e+6 & 2e+6 & 2e+6 & 2e+6 & 2e+6       \\
$k_2$ (g/cm/s$^2$)   & -25 & -35  & -25  & -35  & -25  & -35  & -25  & -35        \\
$k_3$ (g/cm/s$^2$)  & 3e+5  & 3.04e+5  & 3.6e+5  & 3.6e+5   & 4.2e+5  & 4.3e+5  & 3e+5    & 2e+5       \\
$ks_1$ (g/cm/s$^2$) & 2e+6 & 2e+6  & 2e+6 & 2e+6 & 2e+6 & 2e+6 & 2e+6 & 2e+6        \\
$ks_2$ (g/cm/s$^2$)    & -25 & -35  & -25  & -35  & -25  & -35  & -25  & -35       \\
$ks_3$ (g/cm/s$^2$) & 2e+5  & 3.04e+5  & 2.4e+5  & 3.6e+5   & 2.8e+5  & 4.3e+5  & 3e+5    & 3e+5       \\
$p_0$ (mmHg) & 7.7   & 75    & 7.7  & 75 & 7.7 & 75 & 7.7 & 75         \\
$T$ (s)  & 0.615  & 0.615   & 0.615 & 0.615 & 0.615  & 0.615  & 0.658  & 0.658      \\
$r_{\text{min}}$ (cm)  & 0.01 & 0.01 & 0.005 & 0.005  & 0.005  & 0.005  & 0.01  & 0.01       \\ \hline
\end{tabular}
\label{table:parameters}
\end{table}
\begin{table}[b]
\centering
\caption{Global parameter values for all patients. {$\rho$ denotes density, $\mu$ is viscosity, $\alpha$ and $\beta$ are structured tree parameters determining the asymmetry of the tree, and $lrr$ is the length-to-radius ratio of the structured tree.}}
\begin{tabular}{lr}
\hline
Parameter         & \multicolumn{1}{l}{Value} \\ \hline
$\rho$ (g/cm$^3$) & 1.057                     \\
$\mu$ (g/cm/s)    & 0.032                     \\
$\alpha$          & 0.90                      \\
$\beta$           & 0.60                      \\
$lrr$             & 50                        \\ \hline
\end{tabular}

\label{table:CommonValues}
\end{table}

\subsection{Perfusion}
The perfusion and pressure distribution is determined by mapping the 1D flow and pressure predictions from the vessels' terminals to the liver volume \cite{Colebank_perf}. The end of each terminal vessel is denoted by the $x,y,z$ coordinate extracted from the centerlines. Each of these points is assigned the mean flow and pressure from the vessel in question. The flow and pressure are distributed to the volume using Matlab's nearest-neighbor algorithm, providing a 3D representation of blood perfusion and pressure distribution in the liver. To accurately compare differences between patients with different inflows, the flow at the terminals of the hepatic and portal networks are normalized by the average inlet flow of the DORV patient.

To quantify the differences in distributions between patient types, the Kullback-Leibler (KL) divergence is computed \ref{Table:KLdiv}. The KL divergence measures the relative entropy between two probability distributions. We construct both flow and pressure probability distribution functions (PDFs) and determine the following:
\begin{equation}
    D_{KL}(A|B) = \sum_{\omega\in W} A(\omega)\log(\frac{B(\omega)}{A(\omega)},
    \label{eq:KL}
\end{equation}
where $A$ is the ``true'' distribution. We assume the DORV flow and pressure PDFs to be true as they have not undergone aortic reconstruction, and to compare degrees of FALD, we let the HLHS flow and pressure PDFs be the true distribution. $B$ is the distribution of the HLHS patient, the patient with early FALD, and the patient with severe (late) FALD. $A$ and $B$ are defined on some probability space, $W$. Identical distributions can be inferred if $D_{KL}\rightarrow0$. Increasing values of $D_{KL}$ indicate a greater mismatch between $A$ and $B$.

\subsection{Simulations}
\label{sec:sim}
\begin{figure}
    \centering
    \includegraphics[width=1.0\linewidth]{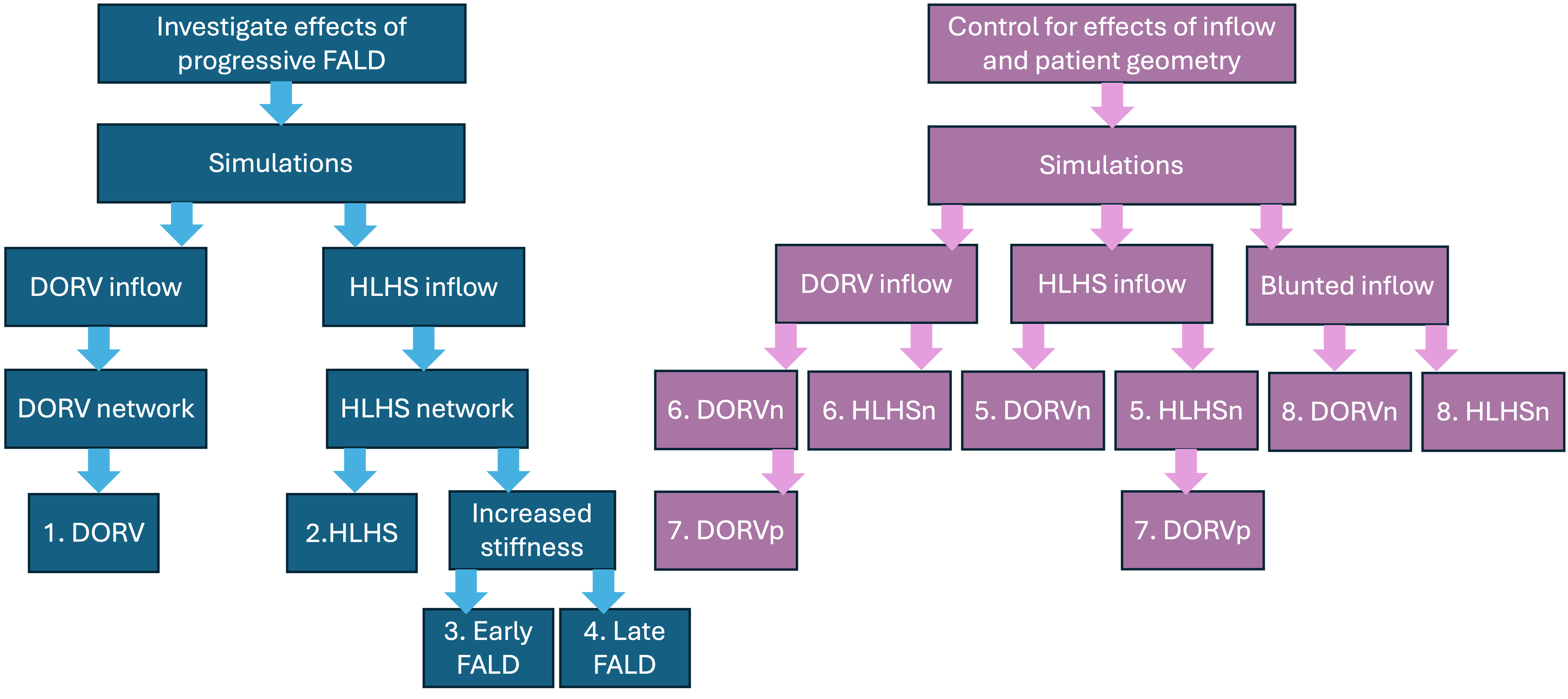}
    \caption{ {Workflow chart. The number (1-8) corresponds to simulations described in \ref{sec:sim}. DORVn refers to the DORV patient network geometry, HLHSn refers to the HLHS patient network geomtery, and DORVp refers to the DORV parameters listed in Table \ref{table:parameters}. Figures of results for simulations 5-8 can be found in the supplement.}}
    \label{fig:placeholder}
\end{figure}
We simulate liver hemodynamics for the DORV and HLHS networks. To mimic FALD in the HLHS patient, we increase the vascular stiffness in the large and small vessels. For each network, the inflow propagates downstream through the large vessels of the portal vein and hepatic artery networks. We define four patient types:
\begin{enumerate}
    \item A control DORV patient based on patient data.
    \item A HLHS patient based on patient data.
    \item An HLHS patient with early FALD progression.
    \item An HLHS patient with late FALD progression.
\end{enumerate}

 {To study the impact of inflow waveform, network geometry, and model parameters, we also include the following simulations (results shown in supplement):}

\begin{enumerate}
    \item [5.]  {\textbf{HLHS inflow waveform effects}: We imposed the HLHS inflow on the DORV and the HLHS networks, respectively,  keeping the vascular parameters and patient specific geometry fixed,}
    \item [6.]  {\textbf{DORV inflow waveform effects}: We imposed the DORV inflow on the DORV and the HLHS networks, respectively,  keeping the vascular parameters and patient specific geometry fixed,}
    \item [7.]   {\textbf{Parameter effects}: Fixing the inflow and network geometry we imposed the DORV parameters on the DORV and HLHS networks.}
    \item [8.]  {\textbf{Portal vein Inflow waveform amplitude}: As reported by \cite{chavan}, the inflow waveform to the portal vein is often blunted. To study effects of this waveform shape, we imposed  digitized the waveform from \cite{chavan} scaling it to the means for the two patients in this study. These simulations were conducted with the patient parameters reported in Table \ref{table:parameters}. }
\end{enumerate}

 {Results for simulations 5-8 are reported in the supplemental material.}

\subsubsection{HLHS vs. DORV} 
To predict the liver hemodynamics in each patient, we calibrate the models using the dimensions in Table \ref{table:parameters} and Table \ref{table:CommonValues}. {We manually tune the nominal stiffness parameters of the large vessel for each patient to generate flow waveforms and pressure predictions that give physically meaningful results, with average pressures and flows being comparable to literature values} \cite{olufsen2000numerical, garcia-pagan1996physical}. 

In this study, the stiffness of the small vessels, ${k}_\text{sn}$, is set equal to their counterparts of the large vessels, ${k}_\text{n}$ \cite{alyssa2022, olufsen2000numerical}.

\subsubsection{FALD Simulation} 
Patients with liver cirrhosis experience increased vessel stiffness due to alterations in the composition of the extracellular matrix (ECM) \cite{arriazu2014extracellular, villela2016NAFLD}. To predict the change in hemodynamics in the diseased patient under the development of FALD, we simulate a liver diseased with FALD by increasing downstream vascular resistance. We increase the parameter $fs_3$, resulting in a stiffness of 3.0 x 10 $^6$(g/cm/s$^2$) in the portal vein and 1.0 x 10 $^6$(g/cm/s$^2$) in the hepatic artery. Furthermore, we decrease the outlet boundary condition $r_min$ to $0.005$ cm to further increase the resistance. 

\section{Results}

We compare the flow and pressure predictions of a DORV, HLHS, and simulated HLHS patient with FALD to investigate differences in hemodynamics.  {We also investigate the impact of cardiac output and inflow waveform by imposing patient inflows on their counterpart network.} Our results show that the HLHS patient has higher portal vein pressures than the DORV patient and that pressure reaches hypertensive levels under FALD conditions.  {These same trends are seen regardless of the inflow waveform imposed.} We see that WSS is higher in the HLHS patient and increases with FALD. The perfusion plots show that the DORV patient has more evenly distributed flow throughout the tissue. It also shows a change in the flow distribution as FALD conditions increase. 

\subsection{Pressure and flow}
\begin{figure}[t]
    \centering
    \includegraphics[width=0.9\linewidth]{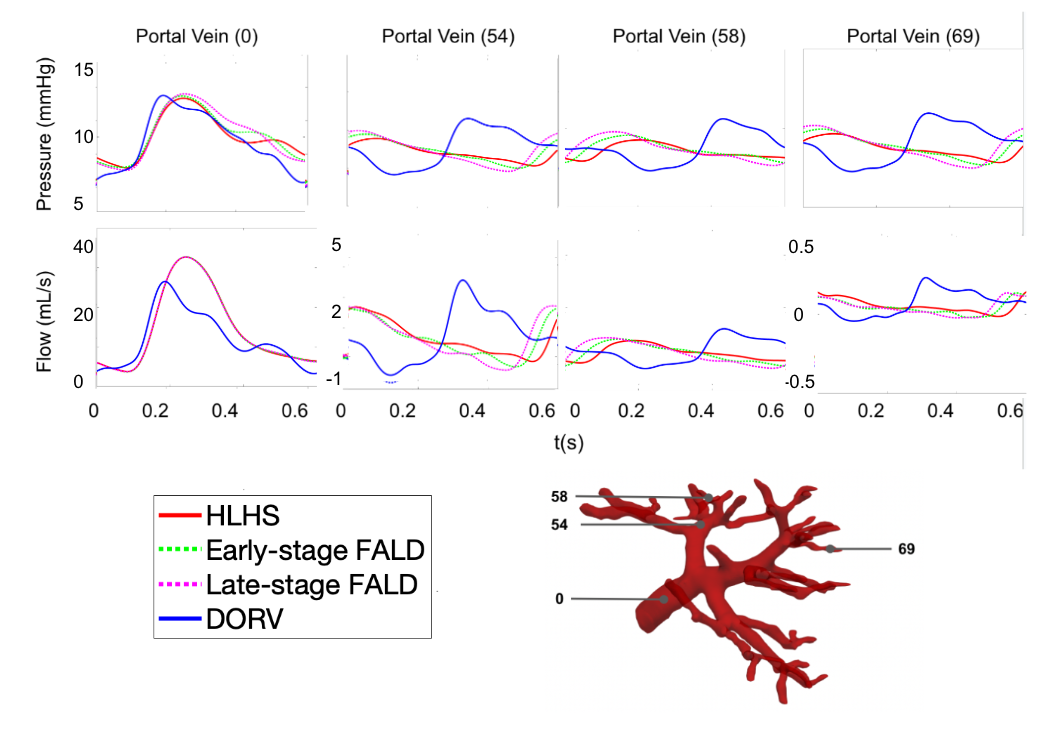}
    \caption{Pressure and flow predictions for the DORV, HLHS, and simulated FALD HLHS patient in the portal network. We see that as the degree of FALD increase, the peak of the flow waveform is slightly shifted to the left, with systolic pressures slightly increasing.}
    \label{fig:portalprediction}
\end{figure}
\subsubsection{Patient-specific inflows}
\noindent\textbf{Portal vein.} 
Table \ref{Table:PortalPressure} shows the predicted pressure and flow in the portal vein network. The average pressure is calculated as the mean pressure during the cardiac cycle. All patient types have approximately the same systolic pressure, with the DORV patient having a lower diastolic pressure. Naturally, the DORV patient also has the highest pulse pressure of the patient groups. However, of all patient types, the DORV patient has the lowest average pressure. In patients with simulated FALD HLHS, systolic pressures and pulse pressures increase slightly. The Late FALD patient has the highest systolic pressure and pulse pressure of the three HLHS patients. Regarding flow, FALD conditions shift the peak of the flow waveform to a somewhat earlier time in the cardiac cycle in the downstream vessels (Figure \ref{fig:portalprediction}).
\begin{table}
\centering
\caption{ {Portal vein pressure predictions for the HLHS and DORV patients. All pressures shown have units mmHg. The Pulse pressure refer to the range between systolic and diastolic pressures. All values are listed in mmHg. Note, Sim 5 refers to the HLHS inflow imposed on DORV network, Sim 6 to the DORV inflow imposed on HLHS network, Sim 7 to the DORV parameters imposed on the HLHS patient network, and Sim 8 to the blunted inflow waveform (a) imposed on the HLHS and (b) imposed on DORV.}}
\begin{tabular}{lrrrrrrrrr}
\hline
  & \multicolumn{1}{c}{HLHS} &
  \multicolumn{1}{c}{Early FALD} &
  \multicolumn{1}{c}{Late FALD} &
  \multicolumn{1}{c}{DORV} & 
  \multicolumn{1}{c}{Sim 5}& 
  \multicolumn{1}{c}{Sim 6}& 
  \multicolumn{1}{c}{Sim 7}&
  \multicolumn{1}{c}{Sim 8a}&
  \multicolumn{1}{c}{Sim 8b}\\
  \hline
Systolic Pressure  & 12.6 & 12.7 & 12.9 & 12.7 & 14.1 & 14.1 & 12.4 & 11.1  & 11.0 \\
Diastolic Pressure & 7.9  & 7.9 & 7.8  & 7.0 & 7.0 & 6.8 & 7.9 & 9.8 & 9.7\\
Pulse             & 4.7  & 4.8  & 5.1  & 5.7 & 7.1 & 7.3 & 4.5 & 1.3 & 1.3 \\ 
Average Pressure   & 10.0  & 10.1 & 10.1 & 9.7 & 10.6 & 10.5 & 10.2 & 10.5 & 10.4\\
%Average Flow (mL/s)       & 15.85  & 15.85 & 15.85 & 11.85 \\ \hline
\end{tabular}
\label{Table:PortalPressure}
\end{table}
To analyze the change in pressure and flow in specific vessels, we selected four representative vessels (0, 54, 58, 69) in the portal vein vasculature for visual comparison. 

The pressure and flow of the inlet vessel (vessel 0) differ significantly between the two patients. The HLHS patient has a smoother inflow curve, but the magnitude and frequency of reflective waves increase with the severity of FALD. The incident pressure wave becomes steeper and the increase in wave speed causes the reflected wave to coincide with the systolic peak; as a result, the pressure wave loses the dicrotic notch as FALD progresses.

\vspace{0.3cm}

\noindent\textbf{Hepatic Artery}
\begin{figure}[htb]
    \centering
    \includegraphics[width=0.9\linewidth]{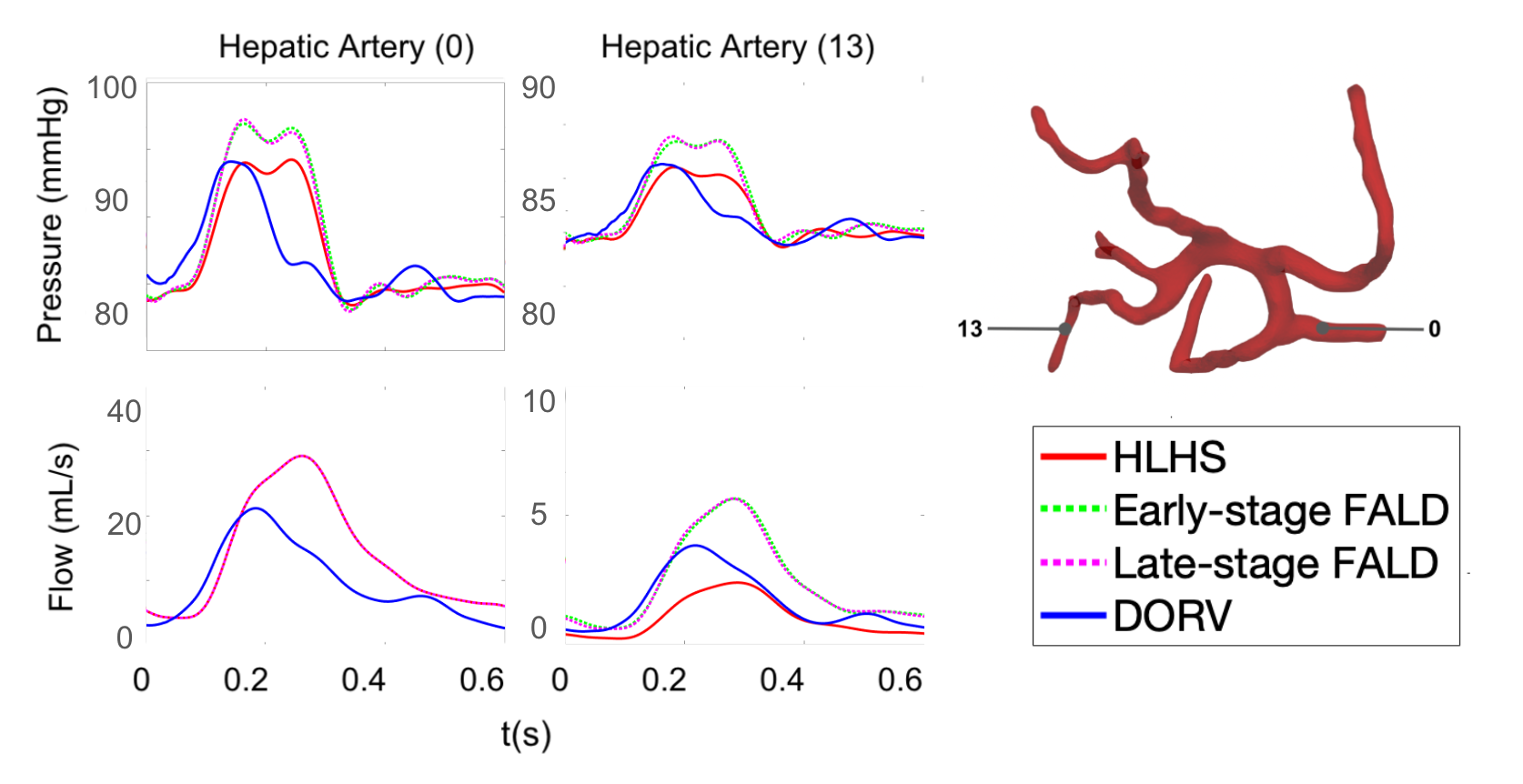}
    \caption{Pressure and flow predictions for vessels 0 and 13 in the hepatic artery vasculature in a DORV, HLHS, and FALD HLHS patient. Systolic pressures are increased under FALD conditions and in this particular terminal vessel (13) the maximum flow increases under FALD conditions.}
    \label{fig:hepaticprediction}
\end{figure}
A similar analysis is performed for the hepatic artery network. Table \ref{Table:HepaticPrediction} shows the predicted pressure and flow in the hepatic artery network. The HLHS patient has a higher systolic and pulse pressure compared to the DORV patient. There is a slight increase in pressures under FALD conditions for the HLHS patient. 

We selected two representative vessels (0, 13) in the vasculature of the liver artery to compare flow and pressure predictions between patients (Figure \ref{fig:hepaticprediction}). Vessel 0 is the hepatic artery, the network's root vessel, while vessel 13 is a terminal vessel. Comparing the healthy HLHS and DORV patients, the HLHS has increased systolic, pulse, and average pressures. Systolic pressures are increased as the degree of FALD increases,with a slight decrease in diastolic pressures. This results in an increase of pulse pressures. 

\begin{table}[t!]
\centering
\caption{ {Predicted pressure and flow in the hepatic artery network of the DORV and HLHS patients. All values are given in mmHg. Sim 5 refers to the HLHS inflow imposed on the DORV network, Sim 6 to the DORV inflow imposed on HLHS network, and Sim 7 to the DORV parameters imposed on the HLHS network.}}
\begin{tabular}{lrrrrrrr}
\hline
  & \multicolumn{1}{c}{HLHS} &
  \multicolumn{1}{c}{Early FALD} &
  \multicolumn{1}{c}{Late FALD} &
  \multicolumn{1}{c}{DORV} &
  \multicolumn{1}{c}{Sim 5} &
  \multicolumn{1}{c}{Sim 6} &
  \multicolumn{1}{c}{Sim 7} \\ 
  \hline
Systolic Pressure  & 95.3 & 99.5 & 99.8 & 95.1 & 89.9 & 104.3 & 97.3 \\
Diastolic Pressure & 83.2  & 82.9 & 82.7  & 83.4 & 83.2 & 83.2 & 83.4 \\
Pulse              & 12.1  & 16.6  & 17.1  & 11.7  & 6.7 & 21.1 & 13.9 \\
Average Pressure   & 88.2  & 88.9 & 88.8 & 88.2 & 86.7 & 93.8 & 90.3\\
%Average Flow (mL/s)       & 6.79  & 6.79 & 6.79 & 5.08 \\ \hline
\end{tabular}
\label{Table:HepaticPrediction}
\end{table}

\subsubsection{Impact of inflow} 
  {Imposing the HLHS (Figure S1) and DORV (Figure S2) inflow on the DORV and HLHS networks, was done to study impacts of the inflow waveform shape. Independent of the inflow waveform results show that simulations with the DORV geometry and parameters (i.e., more compliant vessels) lead to smaller pressure build-up, the pressure is lower in the main hepatic artery (Table \ref{Table:HepaticPrediction}). In the small hepatic artery vessels, the pressures were similar for both patients and inflow. The HLHS inflow is higher, as a result, more flow is pushed through the hepatic arteries, independent of the network. A larger hepatic artery flow results in more flow to the smaller vessels. In the portal network, the higher compliance of the DORV network generates more reflective waves. In this network we observe a significant phase shift in the smaller vessels. Again, similar results were obtained with both inflows.}

  {We also conducted a simulation where the DORV parameterization was imposed on the HLHS geometry to explore impacts of the patient parameterization (Figure S3). Even with the same parameters, similar trends are observed in both flow and pressure predictions (Tables 4 and 5). In the DORV network, the flow maintains a consistent shape and pulsatility, whereas in the HLHS network, the flow becomes flatter in terminal vessels and the pressure are lower in the portal vein network. The main difference is that DORV network has slightly less stiff vessels, as a result wave-propagation speed is slower impacting the arrival time of the reflected waves.}

  {Finally, we imposed a blunted waveform in the portal vein network \cite{chavan} (Figures S4, S5). This waveform shows significant pretense of secondary oscillations resulting from wave reflection generated through the arterial network. Imposing this waveform on the HLHS and DORV network, generates similar flow and pressure amplitudes in both geometries (Table \ref{Table:PortalPressure}). Again, the more compliant DORV vasculature augments secondary flow generating more reflections. }

\subsection{Wall Shear Stress Approximation}
WSS approximations for the hepatic artery and portal vein vasculatures are shown in Table \ref{Table:WSS}. The WSS of the DORV patient is lower in all vessels compared to the HLHS patient.  
\begin{figure}[t!]
    \centering
    \includegraphics[width=0.9\linewidth]{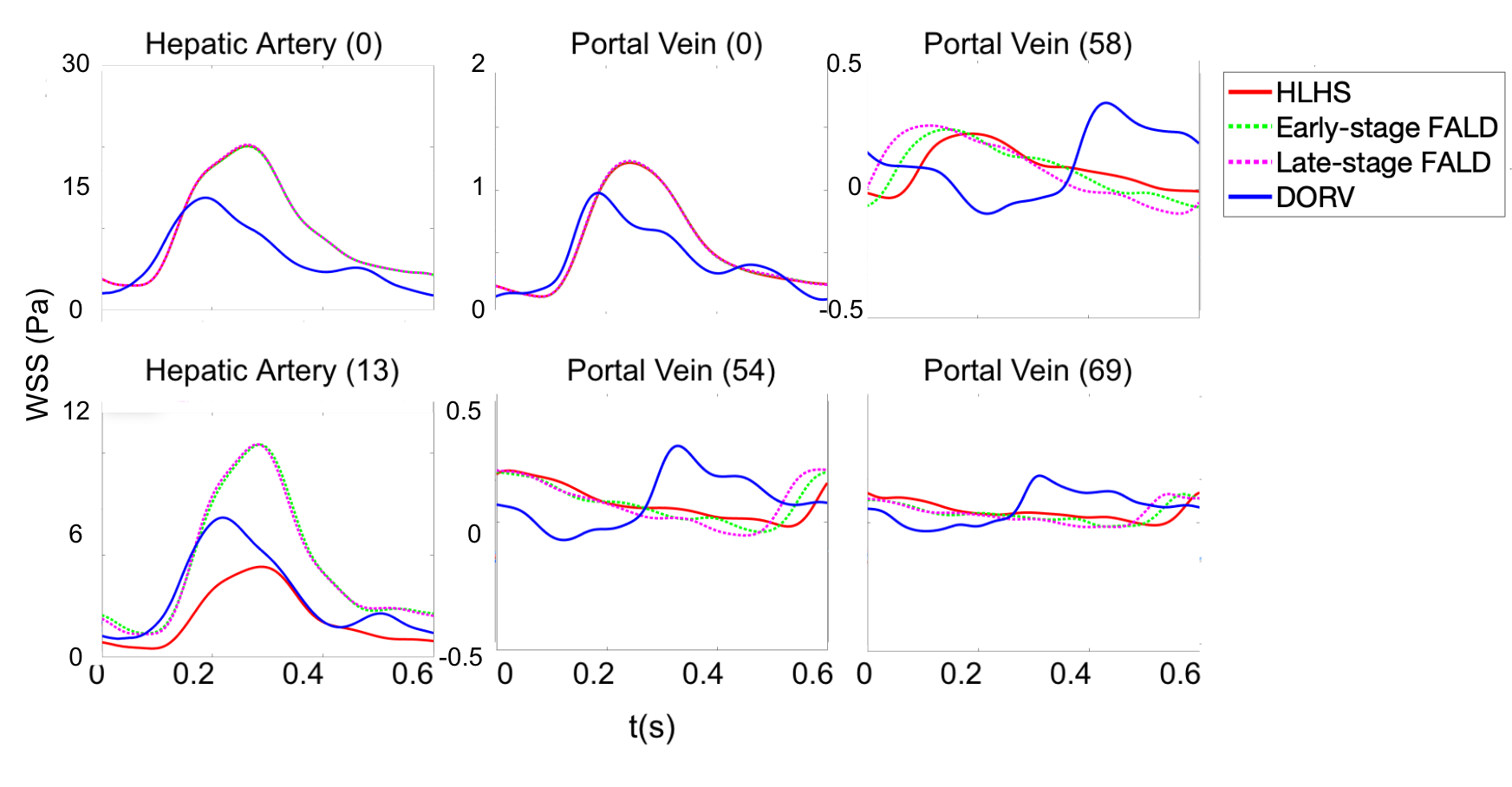}
    \caption{Wall shear stress approximations of hepatic artery vessels (0, 13) and portal vein vessels (0, 54, 58, 69) for HLHS and DORV patients.}
    \label{fig:wss}
\end{figure}

Graphical representations of WSS over one cardiac cycle at the midpoint of the six previously selected vessels are shown in Figure \ref{fig:wss}. Inlet vessels (vessel 0) of both vasculatures show the most significant difference in WSS between HLHS and DORV patients. In the hepatic network, it is clear to see that as FALD progresses, WSS values increase as well. The portal network is not as telling, with only small changes, both increases and decreases, as FALD progresses. The DORV patient tends to have higher WSS values in the downstream portal vasculature.
\begin{table}[b!]
\centering
\caption{Average WSS in the selected DORV and HLHS patient vessels under normal and FALD conditions. Values are given in units of Pa.}
\begin{tabular}{lrrrr}
\hline
                    & Healthy HLHS & Early FALD & Late FALD & DORV  \\ \hline
Hepatic Artery (0)  & 9.61       & 9.60      & 9.65     & 6.50 \\
Hepatic Artery (13) & 1.89       & 4.46      & 4.43     & 2.96 \\
Portal Vein (0)     & 0.05        & 0.06       & 0.06      & 0.04  \\
Portal Vein (54)    & 0.08        & 0.07       & 0.07      & 0.08  \\
Portal Vein (58)    & 0.08        & 0.08       & 0.08      & 0.10  \\
Portal Vein (69)    & 0.04        & 0.04       & 0.04      & 0.6  \\ \hline
\end{tabular}
\label{Table:WSS}
\end{table}

\subsection{Perfusion}
\begin{figure}[H]
    \centering
    \includegraphics[width=1.0\linewidth]{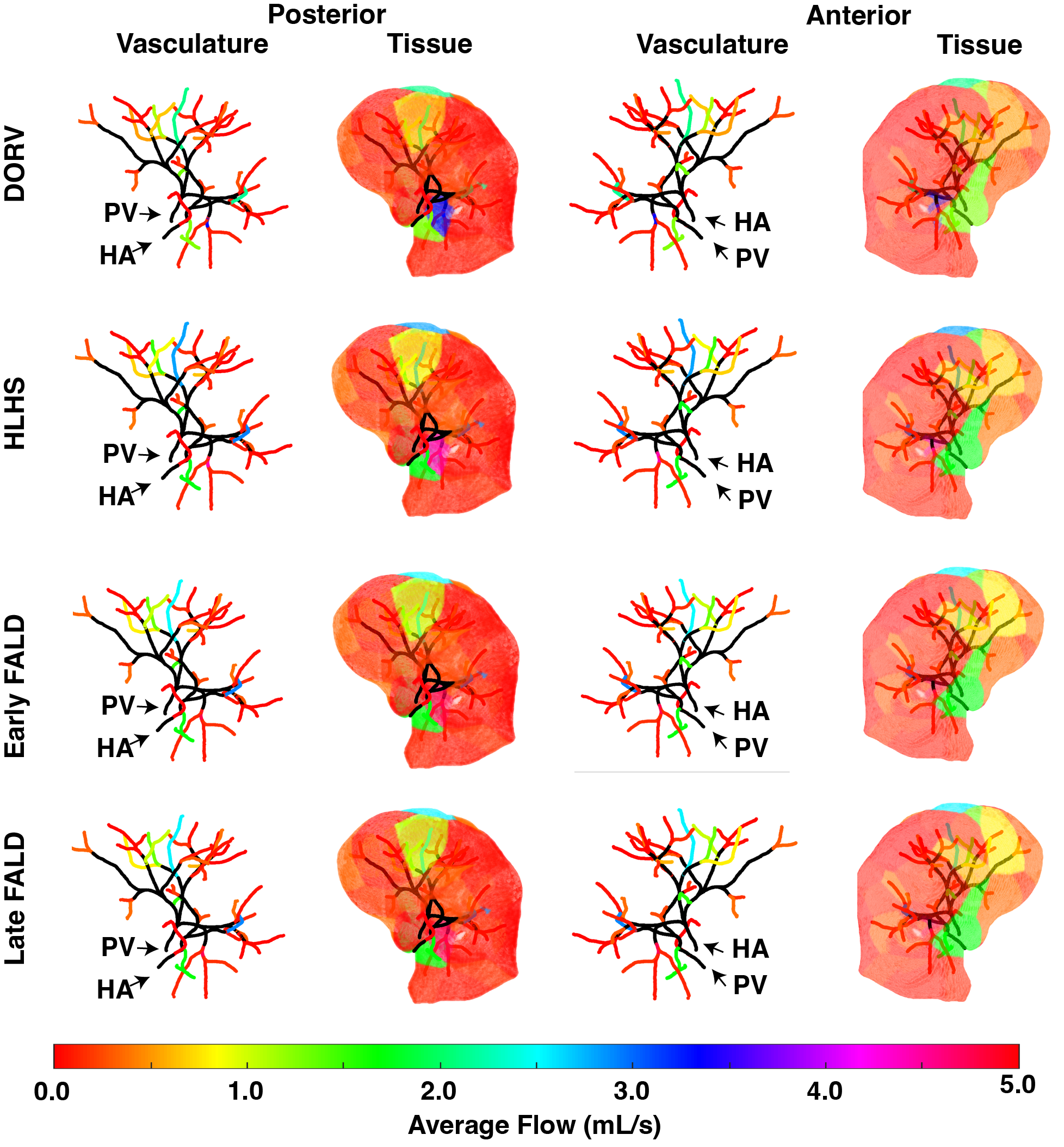}
    \caption{{Perfusion plots distributing the average flow in terminal vessels to the liver. Results are shown at two angles (Posterior and Anterior). For each view, the left column depicts the network and average flow in the terminal vessel, and the right column distributes the flow at the end of each terminal vessel to the nearby tissue. Since the inflow to the HLHS network remains unchanged with disease, the flow distribution does not vary significantly between HLHS cases. In general, the flow decreases slightly as disease severity is increased. Note the light red shown in the anterior view refers to values from 4.8-5.0 mL/s whereas the deep red shown in the posterior view refers to values from 0-0.2 mL/s.}}
    \label{fig:flow_perf}
\end{figure}

\begin{table}[b!]
\centering
\caption{KL divergence values comparing the probability distributions of each HLHS-type patient to the flow and pressure probability distribution of the DORV patient and the distributions of differing degrees of FALD to the healthy HLHS patient. }
\begin{tabular}{l|lrrr}
\hline
 &  & Healthy HLHS & Early FALD & Late FALD \\
  \hline
DORV & Flow  & 124.5 & 38.6 & 146.7 \\
& Pressure & 22.1  & 55.4 & 83.7  \\
\hline
HLHS & Flow  &  & 207.7 & 344.2 \\
& Pressure &  & 37.5 & 76.4  \\
\hline
\end{tabular}
\label{Table:KLdiv}
\end{table}
Perfusion plots (shown in Figure \ref{fig:flow_perf}) are included to visualize the flow distributions within the liver. The plots reveal differences in blood flow to the top and center of the liver. The healthy HLHS patient has the highest amount of perfusion to the top of the liver (light blue color), and as FALD increases, the amount of flow to this region decreases. However, perfusion increases slightly in the posterior and upper part of the liver as FALD increases. Both DORV and HLHS patients have the highest perfusion at the bottom center of the liver (purple region in DORV, pink in HLHS). Overall, the DORV patient has a more evenly distributed flow throughout the tissue. 

KL divergence is calculated to quantify the difference between the flow distributions (Table \ref{Table:KLdiv}). When comparing HLHS patient types to the control DORV, the flow is more similar to the DORV patient in the early FALD HLHS patient. The most significant flow differences are observed in the late FALD HLHS patient. Comparing the healthy HLHS patient with FALD-induced HLHS patients shows that differences are exacerbated as the severity of FALD increases. 

\begin{figure}[H]
    \centering
    \includegraphics[width=1.0\linewidth]{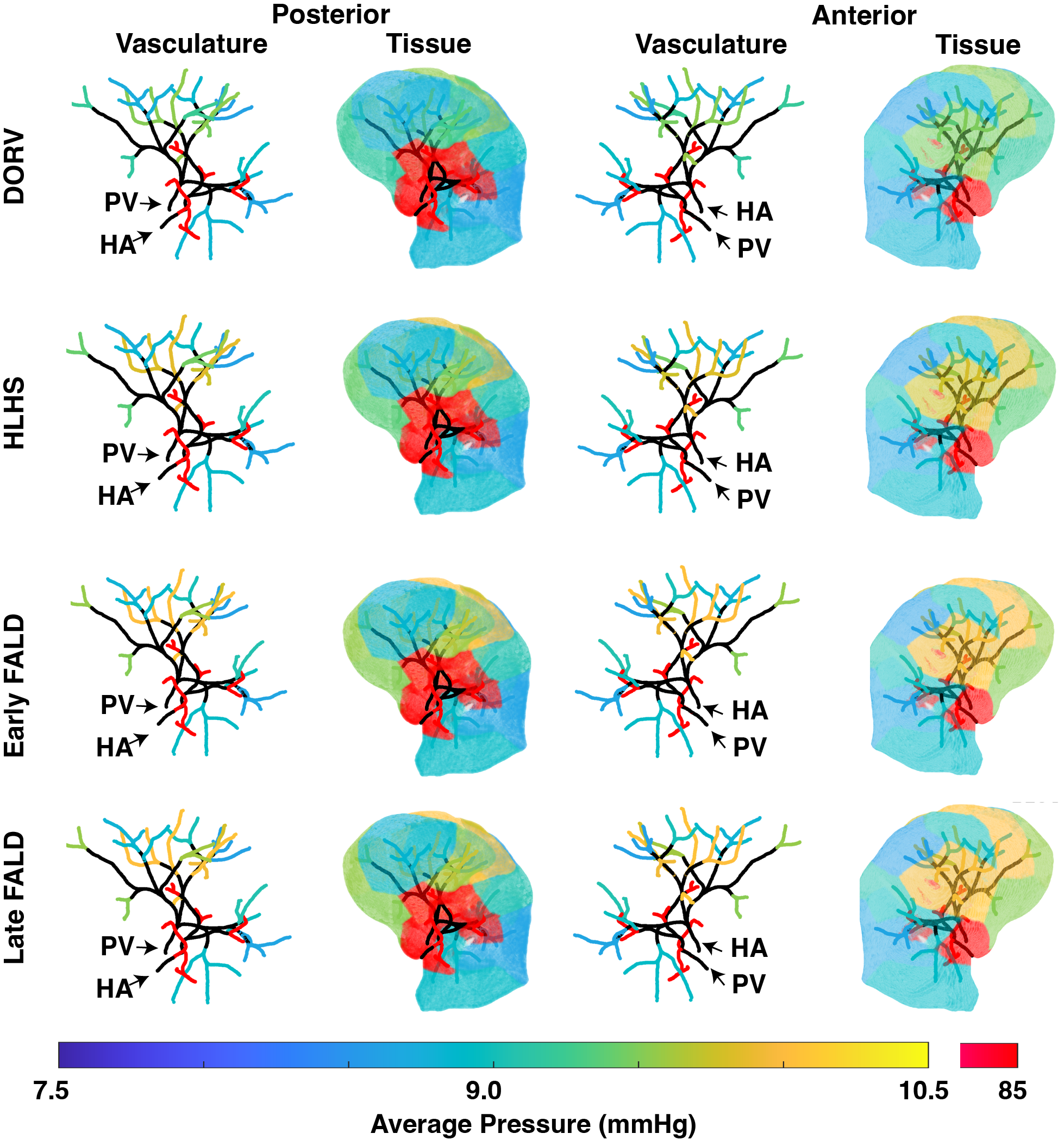}
    \caption{Perfusion plots of the average pressure at terminal vessels in each patient type. Pressures are higher in the portal network in the HLHS patient compared to the DORV. Pressures also increase at the terminals of the portal network as the level of FALD increases.}
    \label{fig:pressure_perf}
\end{figure}

The corresponding pressure distributions are shown in Figure \ref{fig:pressure_perf}. Similarly to flow, results are shown from the posterior and anterior angles. The hepatic artery network transports blood at a significantly higher pressure than the portal vein network. Differences with increased severity of FALD are more pronounced for pressure than flow. There is an increase in pressure in the portal vasculature of patients with HLHS compared to that of the DORV patient. This slight increase suggests that the pulse pressure is also increasing. Increases in pulse pressure, even incremental increases as seen here, indicate that patients may be entering hypertensive levels. Changes in the hepatic network at the terminals are negligible and range from $84.5$ to $85$ mmHg.

KL divergence values (Table \ref{Table:KLdiv}) show that the pressure distribution is similar between the healthy patient with HLHS and the DORV patient, while increasing the severity of FALD leads to a more significant mismatch in the pressure distribution throughout the tissue. The same trend is seen when comparing FALD-induced patients with healthy HLHS patients. As FALD increases, KL divergence values increase, showing more significant differences in pressure distributions.

\section{Discussion}
In this study, we construct a detailed model of the liver vasculature to explore the hemodynamics of patients with HLHS and DORV. We also impose FALD conditions on the HLHS patient to predict pressure and flow under FALD progression.  {The impact of cardiac output and inflow waveform is also investigated by imposing patient inflows on their counterparts. Results show that the HLHS patient has higher pressure than the DORV patient in the portal and hepatic networks. These values reach hypertensive ranges when FALD is imposed. The HLHS network continues to have increased pressures regardless of waveform and parameters imposed.} WSS is also higher for HLHS patients in the hepatic network, also increasing with FALD severity. The increased WSS is an indicator of vascular remodeling and contributes to the development of fibrosis \cite{Eshtehard2017}. The perfusion and pressure distribution plots show the increased pressure changes with increasing levels of FALD, giving insight into regions of tissue that will be affected by vascular remodeling and fibrosis development due to this increase in pressure.

\subsection{Flow and Pressure}
Pressure and flow predictions are generated in both hepatic and portal networks. This is a unique contribution of this study in that it considers both inlet networks. The patient parameters are chosen based on a study by Taylor-LaPole \textit{et al.} \cite{alyssa2022}, and the patient input waveforms are generated using data from the above research.

Model predictions produce a higher average hepatic artery pressure for the HLHS patient, as shown in Table \ref{Table:HepaticPrediction}. Although there is only slight variation in the average pressure, there is a significant increase in the predicted systolic and pulse pressure when FALD is imposed, and these values enter potentially hypertensive conditions \cite{banerjee2012portalhypertension}.  {This increased pressure is observed in the hepatic network, regardless of inflow or parameters imposed.} Progression of FALD also causes and increase in pulse pressure in the portal network. The increase in pulse pressure is an indicator of increased vascular stiffening \cite{MayoClinicPulsePressure}. Portal hypertension and vascular stiffening is a key characteristic of FALD, often associated with the development of fibrosis \cite{emamaullee2020fontan}. Although many patients do not present outward symptoms of FALD until an irreversible stage (i.e., organ failure, liver transplant required), the model is designed to capture the degree of hypertension within Fontan patients. Hypertension indicates vasculature remodeling, fibrotic development, and venous congestion. It is considered a multiorgan disorder associated with abnormal hemodynamics in the systemic and portal vasculatures \cite{Macmathuna2012}.  {Both DORV and HLHS portal pressures increases when their counterpart's inflow was imposed on their network (Table \ref{Table:PortalPressure}). This could be due to the larger impact of inflow on vessels of smaller radius which is expected by Poiseuille's law \cite{Poiseuillelaw}.}
%The increase in pressure for FALD conditions in the hepatic artery. This is reflective of the hepatic arterial buffer response (HABR), an intrinsic regulation of hepatic arterial blood flow \cite{Lautt1985}. Our result agrees with the study by \cite{Gulberg2002}, suggesting that HABR is preserved even in patients with advanced cirrhosis with significant portal hypertension. 
It is also important to note that our model predictions for the hepatic network are consistent with those from the full body network of the same DORV and HLHS patients in the study by Taylor-LaPole \textit{et al.} \cite{alyssa2022}. 

\subsection{Wall Shear Stress}
{The HLHS patient has higher WSS values than the DORV patient in the hepatic artery and portal vein, as shown in Table \ref{Table:WSS}. In the hepatic network, WSS increases under FALD conditions. The higher WSS may indicate increased fibrosis and vascular remodeling \cite{Eshtehard2017,Piva2012}. As discussed in \cite{Espina}, WSS values in the hepatic artery can range from 1 to 9.5 Pa depending on the location of the arteries they are obtained in. However, these values normally fall within 3 to 8 in the hepatic artery itself and from 1-3 Pa in the downstream hepatic vessels. All HLHS patients have increased hepatic artery WSS values and those with FALD conditions imposed have increased WSS values in the downstream vasculature. %As discussed in Taylor-LaPole \textit{et al.}, increased WSS in the HLHS patient is not uncommon, as patients with reconstructed aortas tend to have increased WSS values downstream of surgical reconstruction due to vascular remodeling \cite{alyssa2022,voges}.% Our results for the hepatic network can be confirmed by the study by Taylor-LaPole \textit{et al.} \cite{Lapole2024}, where the WSS for single-ventricle patients continues to increase with decreasing radius. 
The results in the portal network closely resemble the results obtained by Bartolo \textit{et al.} \cite{bartolo2022numerical} in the pulmonary vasculature. Although the portal vein is systemic, its pressure values and physiology closely resemble the arteries within the pulmonary vasculature. %A study by Voges \textit{et al.} \cite{voges} found that patients with reconstructive aortic surgery will have downstream vessels try to dilate in the years following surgery to compensate for the decrease in cardiac output. This dilation increases the WSS, which is consistent with the findings in our analysis. 
In the portal network, the DORV tends to have higher WSS values in the downstream vasculature. Low WSS values in the portal vein tend to indicate portal hypertension \cite{wei2017wall} and portal hypertension is indicative of liver fibrosis \cite{SukKim}. An experimental study by Traub \textit{et al.} \cite{traub} found that low WSS is correlated with the upregulation of vasoconstrictive genes. This upregulation leads to increased vascular modeling, fibrosis, and hypertension \cite{iwakiri2014pathophysiology}. While all the WSS values are low compared to the typical value Wei \textit{et al.} \cite{wei2017wall} defines (1.0 Pa), in downstream vasculature the HLHS patients do have further decreased values compared to the DORV patient.}

\subsection{Perfusion}
The perfusion plots reveal that the DORV patient has a more uniform flow distribution throughout the liver tissue than the HLHS patient. For the healthy patient with HLHS, most blood goes to the upper and lower regions of the liver. As FALD progresses, flow in these regions is reduced while flow increases in the upper-center of the liver (yellow/green region). These differences in blood delivery are confirmed with increasing KL divergence values as FALD conditions increase. This is indicative of vascular remodeling in small downstream vessels \cite{Kolega_remodel}. It is important to note that blood flow changes occur in tissue regions fed by the portal network. This change in flow, most likely the result of vascular remodeling, is a direct response to portal hypertension as the body tries to relieve downstream resistance in regions of the liver that develop fibrosis \cite{Lemoinne2015}. Although this change in flow can reveal indicators of FALD progression, it is not routine practice to monitor blood flow within the liver until patients develop severe outward symptoms. CT and ultrasound are most commonly used to monitor the development of fibrosis \cite{gordonwalker2019fontanliver}. This demonstrates the importance of the mathematical model from which blood distribution in the liver can be detected to provide clinicians with information if more investigation or intervention is required.

The most common indicator of advanced FALD is portal hypertension \cite{gordonwalker2019fontanliver}. The plots showing the pressure distribution demonstrate the pressure differences in specific tissue regions between patient types. The DORV patient has the most consistent pressure distribution throughout the tissue, with very slight differences in the pressures in the tissues fed by the portal network. The pressure in these portal regions increases in the healthy HLHS patient. The left posterior side of the liver has an increase in pressure, and the upper center portion of the anterior side. These regions continue to increase in pressure, reaching hypertensive values as FALD severity increases. As mentioned above, portal hypertension leads to fibrosis. Fibrosis increases downstream resistance, creating a cycle of continual increasing pressures and reduced blood flow \cite{gordonwalker2019fontanliver}. 

\section{Limitations}
Our study is the first to use a 1D-CFD to study the pathology of FALD. The computational efficiency of the 1D model employed in this study facilitates a rapid calibration process with clinical data. Our structural tree allows us to predict flow across the entire portal and the hepatic vascular network, which includes vessels beyond the imaged area. 

Since our model is one-dimensional, we cannot predict the hemodynamics on different sides of the portal vein and hepatic artery walls. This is relevant because sclerosis could occur, especially when the WSS is low in one area of the vessel \cite{zhou2023wallshearstress}. {In future studies, a description of energy loss can be included in the 1D model, as in several previous studies \cite{Colebank21, mynard2015unified}. As Taylor-LaPole \textit{et al.} \cite{alyssa2022} discussed, we can incorporate an energy loss model that could be calibrated by comparing 1D and 3D simulations, for example, by identifying potential areas of secondary flow directly from Four-Dimensional Magnetic Resonance Imaging (4D-MRI) data.} 		

{This study focused on the effects transmitted by the aorta, although clearly the system is connected, and therefore, we should also have added a model of the venous system. Moreover, respiration also affects flow \cite{respiration}, another feature not studied here. Although these factors likely influence the findings, they also add complexity to the system, which, without more data, can be challenging to validate.}

{Another limitation is the network itself.} We have limited our study to the liver vasculature, neglecting the upstream network that begins at the heart. As a result, the model does not account for the potential hemodynamic disturbances that would propagate from the reconstructed aorta. We used a prescribed inflow waveform predicted from the study by Taylor-LaPole \textit{et al.} \cite{alyssa2022}. We did not combine the portal vein and hepatic networks, which can cause some information to be lost. {A two-sided model was not incorporated, neglecting hemodynamic effects of the Fontan conduit where some backward flow tends to occur. In future studies, we plan to combine the hepatic artery and portal vein networks with a whole-body, two-sided network containing the patient's aorta and Fontan conduit.} In addition, perfusion analysis was performed by mapping the predictions at the vessels' terminals to segmented liver tissue. {We plan to incorporate a porous media model at the vessel terminals to allow us to account for mixing and drainage of the blood from the two systems.} This will give a more complete picture of what is happening within the tissue. We also plan to do more studies to reduce cardiac output according to the level of FALD. As FALD progresses and portal hypertension increases, patients will experience a further reduced cardiac output. In our study, we kept cardiac output consistent across all HLHS and FALD patients. Insight from our clinical collaborators, along with greater access to data, will make this future study possible.

During calibration, we manually adjusted the stiffness parameters (see Table \ref{table:parameters}) to fit the expected pressure curve and the portal venous pressure of a healthy individual. However, patients with HLHS are likely to have portal vein hypertension. Thus, the actual pressure of the patient with HLHS may be higher than our predicted pressures. To enhance our model calibration's precision, we can also incorporate multiple pressure readings from various vessels or utilize multiple blood pressure cuff measurements. 

\section{Conclusion}
This study uses a patient-specific 1D-CFD model to predict hemodynamics in the portal vein and hepatic artery vasculatures of an HLHS patient with a reconstructed aorta and a DORV patient with a native aorta. Our model was calibrated to the cardiac output and liver inflow data from a previous study on this data set by \cite{alyssa2022} with parameters in Table \ref{table:parameters} agreeing with normal portal and hepatic venous pressures. We projected cardiac output from DORV and HLHS patients onto the liver model extracted from an adult liver CT and used the 1D-CFD model to predict flow and pressure in the liver networks. We simulated FALD conditions in the HLHS patient to predict pressure and flow across various stages of FALD progression. Our results show that the HLHS patient has increased pressure in the hepatic network compared to the DORV patient,  {regardless of inflow or parameters imposed}. These results were more prominent as FALD conditions increased. The approximate WSS was {higher} in the hepatic vessels in simulated FALD patients. However, WSS values were lower in the portal network for all HLHS patients. Perfusion analysis shows that progression of FALD results in increased pressures in large regions of liver tissue, possibly leading to fibrosis development, and reduced flow to upper and lower regions of the tissue. In general, our study offers information on FALD pathology by showing specific changes in the hemodynamics of the Fontan circulation that could cause portal hypertension and complications of FALD. More clinical 4D-MRI studies are recommended to gain deeper insights into the dynamic hemodynamics of FALD and Fontan-associated circulatory systems.

% Make a comment on percent of cardiac output and how the HLHS has much higher CO than DORV

\section*{Acknowledgments}
We thank the following people from the North Carolina School of Science and Math for setting up this collaboration and approving this research. Dr. Sarah Shoemaker, the Director of Research and Innovation, and Dr. Amy Sheck, the Dean of Science for the research opportunities. and program advisor, Bob Gotwals. We thank Justin Weigand, M.D., for providing access to patient data and guidance on its use.

\section*{Conflict of interest}
We have no conflicts of interest.

\bibliography{refs.bib}

\end{document}